\documentclass[twocolumn,showpacs,showkeys,preprintnumbers,amsmath,amssymb]{revtex4}

\usepackage{graphicx}
\usepackage{dcolumn}
\usepackage{bm}

\begin{document}

\preprint{}

\title{Scale-Free and Stable Structures in Complex {\em Ad hoc} networks
}%

\author{Nima Sarshar}

\author{Vwani Roychowdhury}%
 \email{{nima,vwani}@ee.ucla.edu}
\affiliation{Department of Electrical Engineering, University of California, Los Angeles}%

\date{\today}

\begin{abstract}
Unlike the well-studied models of growing networks, where the
dominant dynamics consist of insertions of new nodes and
connections, and rewiring of existing links, we study {\em ad hoc}
networks, where one also has to contend with rapid and random
deletions of existing nodes (and, hence, the associated links). We
first show that dynamics based {\em only} on the well-known
preferential attachments of new nodes {\em do not} lead to a
sufficiently heavy-tailed degree distribution in {\em ad hoc} networks.
In particular, the magnitude of the power-law exponent increases
rapidly (from $3$) with the deletion rate, becoming $\infty$ in the limit of
equal insertion and deletion rates.
We then introduce
a {\em local} and {\em universal} {\em compensatory rewiring}
dynamic, and show that even in the limit of equal insertion and deletion rates
true scale-free structures emerge, where the degree distributions obey
a power-law with a tunable exponent, which can be made arbitrarily close to -2.
These results provide the first-known evidence of emergence of scale-free degree
distributions purely due to dynamics, i.e., in networks of almost constant average size.
The dynamics discovered in this paper can be
used to craft protocols for designing highly dynamic Peer-to-Peer
networks, and also to account for the power-law exponents observed
in existing popular services.

\end{abstract}

\maketitle

\section{\label{sec:level1}Introduction}

Several random protocols (i.e., stochastic rules for adding/deleting nodes
and edges) that lead to the emergence of scale-free
networks have been recently proposed. The underlying dynamics
for almost all of these models follow the principle
of preferential attachment for targeting or
initiating newly created links of the network. The simplest case
is the linear preference for the target node of a new link:
a node is added to the network at each time step, and
the probability that a node $i$ with $k(i,t)$ links at time $t$ ($t>i$)
receives a new link at time $t+1$, is $\propto k(i,t)$. The
 resulting network for this simple model has a power-law degree
distribution with an exponent $|\gamma|=3$. Other variations of this
procedure have also been widely studied
\cite{Dog1,Krap,BA,chung}.

The interesting properties of random power-law networks
appear when the degree exponent
 $|\gamma|<3$. These properties include almost
constant diameter and zero percolation threshold. Moreover,
almost all cases of power-laws observed in real life networks,
which these models ultimately might want to account for, have
exponents less than $3$. Motivated by both these
issues, a few stochastic linking rules resulting
in exponents with magnitude less than $3$ have been introduced.
Examples of such protocols include,  the doubly-preferential attachment scheme for links,
where both the initiator and the target nodes of an edge are chosen
preferentially, as proposed in \cite{Dog3,chung}, and the rewiring scheme
of existing links to
preferential targets as proposed in \cite{BB}.

Most of these random protocols have been motivated by the need to
model \emph{growing} and mostly \textit{rigid} networks, where
nodes and links are gradually added. Examples of such graphs are
the citation and collaboration networks. Once a connection is made
between two nodes in these graphs it is never deleted and also
nodes never leave the network. A second class of networks that has
been studied is where the nodes are stable, but the links could be
deleted. For example, on the WWW one can assume nodes to almost
always remain in the network once created; however, existing links
can easily be deleted, and new links created. In this paper, we
primarily address a third class of networks (first introduced in
\cite{Dog1}), where the nodes
themselves are also \textit{unstable} and \textit{unreliable}, and
in an extreme case, the nodes (and hence all their connections)
might leave the network without prior notice and through
independent decisions. .

 Our motivation for considering such dynamic networks comes, in part,
 from the recent
 interest towards less structured or \textit{ad hoc} distributed
 system designs with peer-to-peer (P2P) content sharing networks as a prime
 example. In an instance of Gnutella, for example, a study \cite{GNUT}
 shows that almost 80\%
 of all nodes log-off within five hours from their log-in. Hence,
 the time scale within which the network assumes its structure is much
 shorter than the time scale within which it grows. A number of
 crawls of these networks show that at least in some regimes
 they follow a power-law. However, a stochastic model that can lead
 to the emergence of such complex networks has not been proposed.  Another
 significant example is the \textit{ad hoc} and mobile
 communication paradigms where each member can provide a
 short-time \textit{unreliable} service and yet a global topological
 structure is to be ensured at all times.

We first use the continuous rate equation approach introduced in
\cite{Dog1} (see Section \ref{II}) to predict the power-law
exponent for stochastic models, where new nodes joining the
network make links preferentially, and existing nodes in the
network are uniformly deleted at a constant rate. Contrary to
previous claims \cite{Dog1}, we show that for such models the
power-law degree distribution of the resulting network has an
exponent $|\gamma|>3$, and that it rapidly approaches $-\infty$ as
the deletion and insertion rates become equal. Thus a network with
even small deletion rates will essentially have characteristics
that are more similar to an exponential degree distribution. In
Section \ref{III}, we introduce our compensatory rewiring
procedure, which is a novel way to exploit the deletion dynamic of
the nodes itself to maintain a scale-free structure. In this
protocol, in addition to the new nodes making preferential
attachments, existing nodes compensate for lost links by
initiating new preferential attachments. In fact, we show that the
exponent of the power-law for the degree distributions of the
resulting networks for any deletion rate, can be tuned as close to
$-2$ as desired. Thus, our results provide a random protocol for
generating scale-free networks even in the limit where the
deletion and addition rates are equal and the network size is
almost constant. To the best of our knowledge, this is the only
procedure resulting in scale-free structures with exponent
arbitrary close to $-2$ while the network size is almost constant.

These results can be applied for both analysis and design of complex networks
(see Section \ref{V}). For example, our results provide
an intuitive account for the existence of scale-free
structures in many of the P2P networks \cite{GNUT}.
Perhaps, more
significantly, {\em our results provide a truly local protocol for generating
highly dynamic scale-free and tunable networks}.
While such scale-free unstructured P2P
networks have been thought  to inherently
suffer from scalability problems related to searching, our
recent results  prove this commonly-held notion to be false,
and show that one can indeed perform searches in highly
scalable fashion on such networks \cite{scalable}.  Thus, one could use
the protocols introduced in Section~\ref{III} to design very active and
efficiently searchable content sharing networks.

\section{Growing networks in the presence of permanent node
deletion} \label{II} The scale free properties of growing networks
that incorporate preferential attachment with {\em permanent deletion
of randomly chosen links} was considered by Dorogovtsev et al
\cite{Dog1}. They concluded that the scale free properties of the
emerging network depends strongly on the deletion rate of the links, and in
fact the scale free behavior is observed only in low deletion
rates. However, the analysis of the effect of {\em random
deletions of nodes at a fixed rate} was incomplete. A correct analysis
is presented in this section, and as noted in the introduction, the
associated results are shown to have far-reaching consequences for ad hoc networks.

\subsection{Preferential attachment and random node deletions}
\label{II-A}
We consider the following model: at each time step, a
node is inserted into the network and it makes $m$  attachments to $m$
preferentially chosen nodes. That is, for each of the links,
a node with degree $k$ is chosen as a
target with probability proportional to $k$. Then with probability
$c$, a randomly chosen node is deleted.

We adopt the same approach as introduced in \cite{Dog1} for our
analysis. Let each node in the network be labelled by the time it
was inserted, and define $k(i,t)$ as the degree of the node
inserted at time $i$ (i.e., the $i^{\hbox{th}}$ node) at time $t$. Let
$D(i,t)$ be the probability that the $i^{\hbox{th}}$ node is not
deleted (i.e., it is still in the network) until time $t$.
Assuming the $i^{th}$ node to be in the network at time $t$, the
rate at which its degree increases is:
\begin{equation}\label{1}
    \frac{\partial k(i,t)}{\partial
    t}=m\frac{k(i,t)}{S(t)}-c\frac{k(i,t)}{N(t)} ,
\end{equation}
where
\begin{equation}\label{2}
    S(t)=\int_{0}^{t}D(i,t)k(i,t)di
\end{equation}
is the sum over the degrees of all nodes \textit{that
are present} in the network at time $t$, and $N(t)=(1-c)t$ is the
total number of nodes in the network. Note that the first term in
Eqn.~(\ref{1}) is
simply the number of links node $i$ receives as a result of the $m$
preferential attachments made by the newly introduced node. The
probability that a randomly chosen node is among the neighbors of node $i$,
and hence the probability that node $i$ loses a link, is of course,
$\frac{k(i,t)}{N(t)}$, which accounts for the second term in Eqn. (\ref{1}).
We neglect all higher order effects.

\begin{figure}
  \includegraphics[width=3in,height=2in]{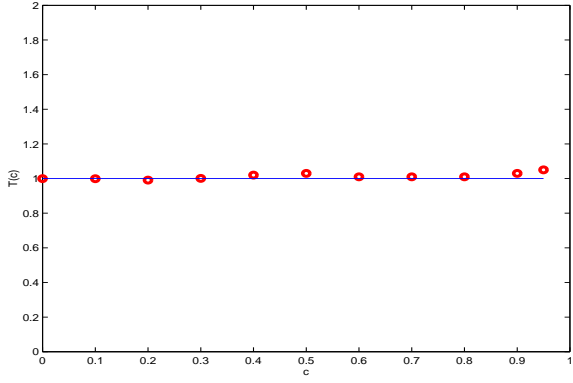}\\
  \caption{$T(c)=\left\langle k\right\rangle \frac{(1+c)}{2m}$ as a function of $c$, the deletion rate.
  Eqn.~(\ref{6}) predicts this quantity to be $1$ regardless of $c$.
  The simulation is for $m=7$, and $c$ ranges from $0$ to $0.95$, and the simulation
  time is $20000$ time steps.}\label{St-1}
\end{figure}

Next, we solve the various unknown quantities in the
following order: $D(i,t)$, $S(t)$, and then $k(i,t)$. First, using
independence of the events corresponding to random deletions of nodes at each
time step, it is easy to verify that $D(i,t+1)=D(i,t)(1-\frac{c}{N(t)})$. Hence, the
continuous version of the dynamic of $D(i,t)$ can be stated as follows:
\begin{eqnarray}
    \frac{\partial D(i,t)}{\partial
    t} &=& -c\frac{D(i,t)}{N(t)}=-\frac{c}{1-c}\frac{D(i,t)}{t}\ . \nonumber
\end{eqnarray}
Since, $D(t,t)=1$, we get
\begin{equation} \label{3}
D(i,t)=\left(\frac{t}{i}\right)^{\frac{c}{c-1}}.
\end{equation}
To find $S(t)$, we first multiply both sides of Eqn. (\ref{1})
by $D(i,t)$ and integrate out $i$ from $0$ to $t$. Then,
\begin{eqnarray}\label{4}
    \int_{0}^{t}D(i,t)\frac{\partial k(i,t)}{\partial
    t}di=m-c\frac{S(t)}{(1-c)t} .
\end{eqnarray}
The left-hand-side of the above equation can now be simplified as follows:
\begin{eqnarray}\label{5}
    \int_{0}^{t}\frac{\partial}{\partial
    t}\{D(i,t)k(i,t)\}di - \int_{0}^{t}k(i,t)\frac{\partial}{\partial
    t}D(i,t)di & = & \mbox{} \nonumber \\
   &\hbox{\hspace*{-2.75in}} & \hbox{\hspace*{-2.5in}}\frac{\partial}{\partial
    t}\left[ \int_{0}^{t}\{D(i,t)k(i,t)\}di\right] -k(t,t)D(t,t)-\nonumber\\
    & \hbox{\hspace*{-2.75in}}& \hbox{\hspace*{-2.0in}}\int_{0}^{t}k(i,t)\frac{c}{t(c-1)}D(i,t)di\ .
\end{eqnarray}
Substituting the above expression in Eqn.~(\ref{4}), and noting that
$k(t,t)=m$ and $D(t,t)=1$, we get
\begin{equation}
    \frac{\partial S(t)}{\partial t}-m-\frac{c}{(c-1)t}S(t)=m-c\frac{S(t)}{(1-c)t}\ .
\end{equation}
The solution to the above equation is:
\begin{equation}\label{6}
    S(t)=2m\frac{1-c}{1+c}t= 2m\frac{N(t)}{1+c}\ .
\end{equation}
Since the correctness of this last equation is the key to further
derivations and since this equation marks the departure from
the results in \cite{Dog1}, we have paid especial attention to it.
In particular, if we define the average degree of nodes at time
$t$ as $\left\langle k(t)\right\rangle= \frac{S(t)}{N(t)}$, then
Eqn. (\ref{6}) implies that
$$\left\langle k(t)\right\rangle=\frac{2m}{(1+c)}=\left\langle k \right\rangle\ ,$$
i.e., the average degree of nodes is modified by a factor of $(1+c)$.
Fig.~\ref{St-1} depicts the
simulation results verifying the prediction of Eqn. (\ref{6}).

\begin{figure}
  \includegraphics[width=3in,height=2in]{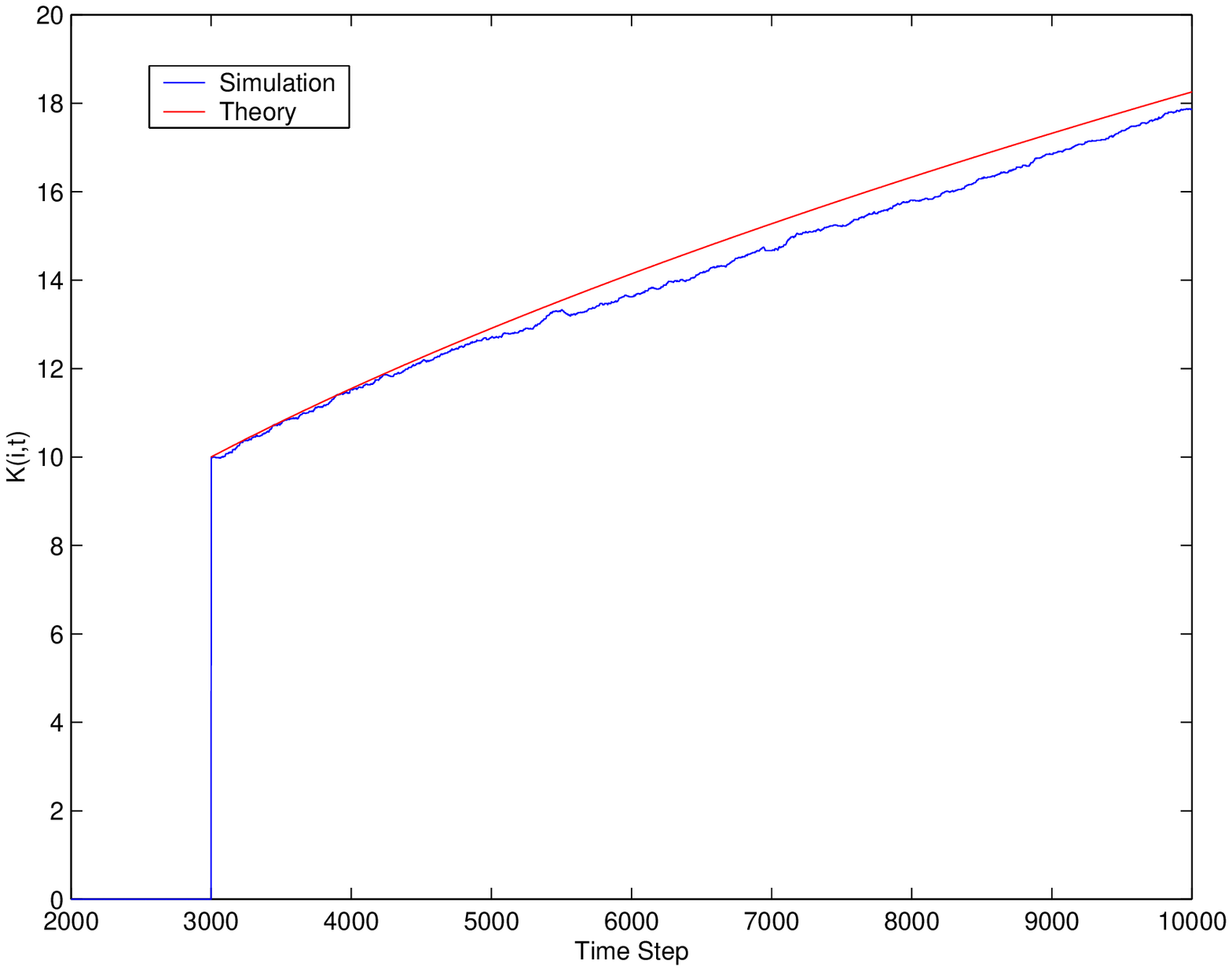}\\
  \includegraphics[width=3in,height=2in]{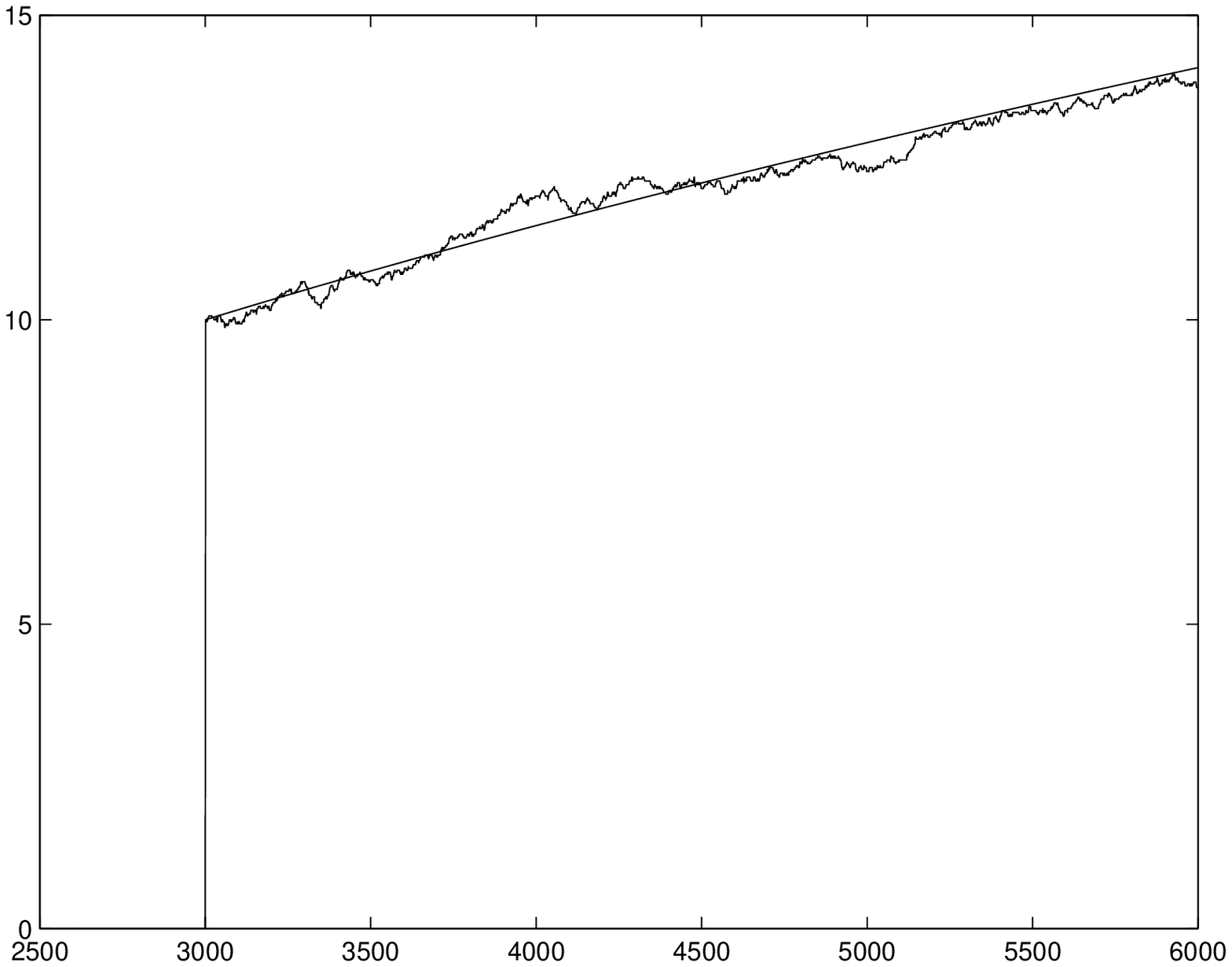}\\
  \caption{The evolution of the degree of a node inserted in the network.
  The power-law growth, and its independence from the deletion rate,
   is at the heart of the results of Section \ref{II}.
  The top figure is the plot for the case of 20\% deletion rate.
  The bottom figure is for the case of 70\% deletion rate.
  A node is inserted at time step $t=3000$, and its degree is recorded at
  future time steps (for $m=10$) until it gets deleted.
  Over 100 trials, the degree of this node (for the trials where it
  was not deleted until time step 10000 and 6000, respectively) are
  averaged and the results are compared to predictions.
  }\label{GROWTH}
\end{figure}

Inserting Eqn. (\ref{6}) back into the rate equation, we get:
\begin{eqnarray}
    \frac{\partial k(i,t)}{\partial
    t}&=& m\frac{(1+c)k(i,t)}{2m(1-c)t}-\frac{c}{1-c}\frac{k(i,t)}{t}\nonumber\\
    &=& \frac{(1+c-2c)k(i,t)}{2(1-c)t}=\frac{k(i,t)}{2t}\ ,\nonumber
    \end{eqnarray}
which implies that
\begin{eqnarray}\label{7}
    k(i,t) &=& m(\frac{t}{i})^{\beta}\ ,
\end{eqnarray}
where $\beta = 1/2$. Eqn.~(\ref{7}) is quite significant
since it states that
the degree of a node in the network (when it is not deleted), does
not depend on the deletion rate. To verify this, we have made
numerous simulations for a wide range of deletion rates. Fig.~\ref{GROWTH}
 shows the results for two rather extreme cases of
20\% and 70\% deletion rates, respectively.

\begin{figure}
  \includegraphics[width=3in,height=2in]{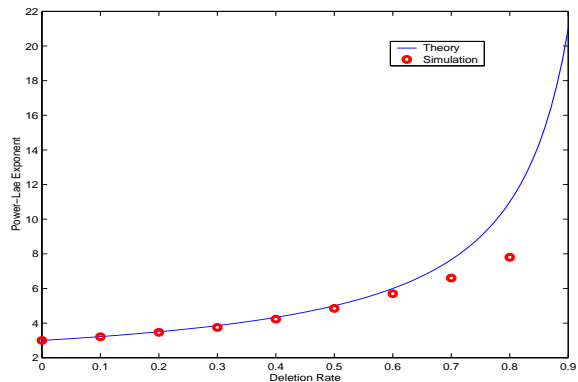}\\
  \caption{The power-law exponent for the degree distribution of networks
  generated with the model discussed in Section~\ref{II}.
  The exponents are the slopes of the best fit line to the log-log plots of
  the corresponding cumulative distributions.
  The time steps at which snapshots are taken vary from 20000 to
  100000 based on the deletion
  rate, so that at the time of snapshot, almost 20000 nodes are in the network
  for all cases. The theory and the simulation results are in
  perfect agreement for small values of $c$.
  For larger values of $c$ however, tracking the very fast growing
  exponent is rather hard. The deviation then seems to be the
  result of the finite number of time steps. In any case, the value of the
  exponent for $c>50\%$ is too large for the network to display any of
  the desirable properties usually associated with scale-free networks.}\label{power-law-c}
\end{figure}

Now to calculate the power-law exponent, we note that
\begin{eqnarray}\label{8}
    P(k,t)& = & \frac{\hbox{No. of nodes with degree $=\ k$}}{\hbox{Total number of nodes}}
    \nonumber \\
    &=& \frac{1}{N(t)}\sum_{i:k(i,t)=k}
    D(i,t)\nonumber \\
    &=&\displaystyle \frac{1}{N(t)}D(i,t)\left\vert\frac{\partial k(i,t)}{\partial
    i}\right\vert_{i:k(i,t)=k}^{-1}
\end{eqnarray}
From Eqn.~(\ref{7}), we obtain:
\begin{eqnarray}
    \frac{t}{i_{k}}&=& m^{-1/\beta}k^{1/\beta}\ , \nonumber
\end{eqnarray}
and thus,
\begin{eqnarray}    \label{9}
    \displaystyle \frac{\partial i}{\partial
    k}\vert_{i=i_{k}}&=& m^{1/\beta}k^{-1/\beta-1}(-1/\beta)t \ .
\end{eqnarray}
Inserting it in Eqn.~(\ref{8}), we get
\begin{eqnarray}\label{10}
    P(k,t)&=&\frac{k^{-\frac{c}{(1-c)\beta}}}{(1-c)m^{\frac{-1}{\beta(1-c)}}}k^{-1/\beta-1}\nonumber \\
    &=&\frac{k^{-1-\frac{1}{(1-c)\beta}}}{(1-c)m^{\frac{-1}{\beta(1-c)}}}\ ,
\end{eqnarray}
which is a power-law distribution with the exponent
\begin{equation}\label{gam}
\gamma=-1-\frac{1}{(1-c)\beta}
\end{equation}
This equation for obtaining the power-law exponent
from Eqn.~(\ref{7}) for a general $\beta$ will be used later
on too. For our case of $\beta=1/2$ we get the exponent of
\begin{equation}\label{gam-c}
    |\gamma|=1+\frac{2}{1-c}\ .
\end{equation}
As illustrated in Fig.~\ref{power-law-c}, simulation results provide a
verification of Eqn.~(\ref{gam-c}).

\subsection{Additional preferentially targeted links will not help}
\label{II-B}
 We now show that introducing new preferential attachments, as introduced in \cite{BB},
 will not help control the divergence of the exponent.
 To see this, let us modify  the protocol as follows:
  At each time step, a new node is added and it makes
$m$ preferential attachments; $c$ randomly chosen links are
deleted; and a randomly chosen node initiates $b$ preferentially
targeted links.

Following the same steps, as in the previous section, one can show
that $\displaystyle S(t)=\frac{2m(1+c)(b+1)t}{1-c},$ and one can
verify that the rate equation would simplify to $\displaystyle
\frac{\partial k(i,t)}{\partial
    t}=
    m\frac{(1+c)(b+1)k(i,t)}{2m(b+1)(1-c)t}-\frac{c}{1-c}\frac{k(i,t)}{t},$
    which is equivalent to Eqn.~(\ref{7}), and results in the same power-law
exponent as in Eqn.~(\ref{gam-c}).

\subsection{The expected degree of any particular node}
The degree of an existing node
is governed by Eqn. (\ref{7}) until it gets deleted, when  its
degree can be assumed to be $0$. Thus, the
expected degree of the $i^{\hbox{th}}$ node at time $t$ is
given by (see \cite{Dog1})
\begin{eqnarray}\label{KD}
    E(i,t)=K(i,t)D(i,t) \nonumber \\
    \displaystyle =m\left(\frac{t}{i}\right)^{-\frac{c}{1-c}+\beta}
    \nonumber \\
    \displaystyle =m\left( \frac{t}{i}\right)^{\frac{-(\beta+1)c+\beta}{1-c}}\ .
\end{eqnarray}
Hence, if we define $\displaystyle c_{0} = \frac{\beta}{\beta+1}$, then
for $c > c_{0}$,  $E(i,t)\rightarrow \infty$, and for $c < c_{0}$,
$E(i,t)\rightarrow 0$ when $\displaystyle \frac{t}{i}\rightarrow \infty$.

\begin{figure}
  \includegraphics[width=3in,height=2in]{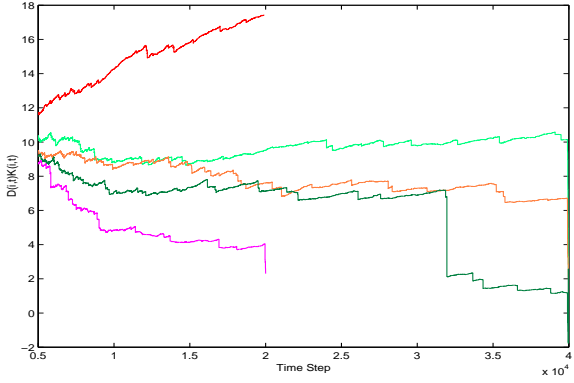}\\
  \caption{A node is inserted at time step=3000,
  and its degree is recorded at future time steps. After it is deleted,
  the degree would be zero for the rest of the time.
  This quantity is averaged over 100 trials of the process to find $E(i,t)$ defined in
  (Eqn. \ref{KD}). If $c>1/3$, then $E(i,t)$ should go
  to zero as $i/t\rightarrow \infty$, otherwise it should go to infinity.
  The curves from top correspond to $c=20\%,30\%,35\%,40\%,50\%$, respectively.}\label{KDof}
\end{figure}

For our case of $\beta=1/2$, $c_{0}=1/3$. Hence, for high enough
deletion rates, a node is not expected to acquire infinite links
before it is deleted. Fig.~\ref{KDof} shows simulation results for
the average degree of nodes and the change in the behavior of
$E(i,t)$ around $c=0.33$ is again in perfect match with our
predictions.

\section{The Compensation process}\label{III}

We now introduce a local and universal random protocol that will
lead to the emergence of true scale-free networks when nodes are
deleted at a fixed rate.
\subsection{Deletion-Compensation Protocol}
 Consider the following process, where at each time step:
\begin{enumerate}
\item  A new node is inserted and it makes $m$
connections to $m$ preferentially chosen nodes.
 \item With probability $c$, a uniformly
chosen node and all its links are deleted.
 \item If a node looses a
link, then to compensate for the lost link it initiates
$n<n_{\hbox{\small crit}}(c)$ ($n$ is real) links, the targets of
which are chosen preferentially. The upper-bound, $n_{\hbox{\small
crit}}(c)$, is specified later.
\end{enumerate}
This protocol is simple in its description as well as in
implementation. It is also truly local, i.e., the decisions for all
nodes (whether to be deleted or to initiate a compensatory link)
are independent and based on the node's own state.

\subsection{Properties of the emergent network}

\subsubsection{Degree distribution}
 The rate equation formulation for the degree of the
 $i^{th}$ node at time $t$ can be stated as:
\begin{equation}\label{11}
     \frac{\partial k(i,t)}{\partial
    t}=m\frac{k(i,t)}{S(t)}-c\frac{k(i,t)}{N(t)}+nc\frac{k(i,t)}{N(t)}+
    nc\left\langle k(t)\right\rangle \frac{k(i,t)}{S(t)}\ ,
\end{equation}
where (I) $S(t)$ is the sum of the degrees of all the nodes
in the network at time $t$ (as defined in Eqn. (\ref{2})).
(II) The first two terms on the right-hand-side
are as described in Eqn. (\ref{1}). (III) The third term accounts for the
fact that the $i^{th}$ node initiates $n$ links if it looses one.
(IV) The fourth term (where $\left\langle k(t)\right\rangle$ is the average
degree of a node) represents the preferential links made to the
$i^{th}$ node by other nodes that lost links because of the deletion
of a uniformly chosen node. The average degree of a node
$\left\langle k(t)\right\rangle = \frac{S(t)}{N(t)}$.

\begin{figure}
  \includegraphics[width=3in,height=2in]{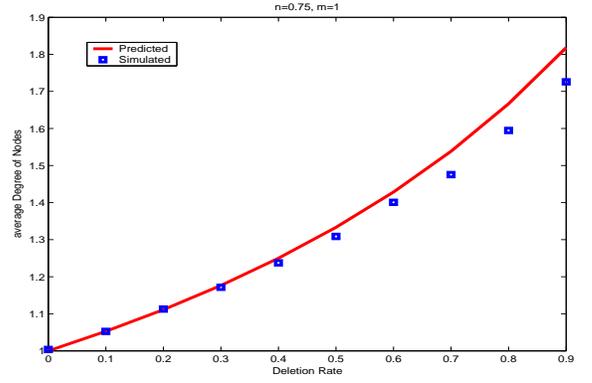}\\
  \caption{Eqn. (\ref{13}), states that: $\left\langle k \right\rangle_c=\frac{m}{1+c-2cn}$.
  For the case $n=0.75$, the value $\left\langle k \right\rangle_c$ is compared with the simulated
  results for $0\leq c \leq 0.9$ and $m=1$. The
  simulations are in good agreement with the predictions, especially for values of
  $c$ less than $0.7$.}\label{GROWTH2}
\end{figure}

\begin{figure}
  \includegraphics[width=3in,height=2in]{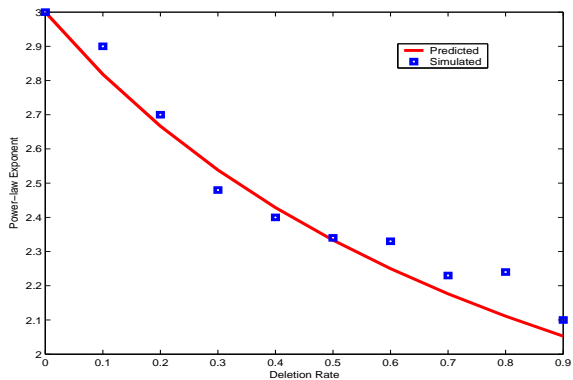}\\
  \caption{The power-law exponent for different values of the deletion rate.
  For all values of $c$, the number of nodes at times the snapshots were taken,
  were kept to be at least 20000.
  The value of the power-law exponent is the
  best fit to the histogram of the cumulative  data. No fit had regression
  confidence less than 98.3\% for the log-log case . The theoretical curve
  is depicted in solid. The simulation results are
  indicated by $\square$.}\label{pl-exp-comp}
\end{figure}

Note that $D(i,t)$ is still given by Eqn.~(\ref{3}). Next,
instead of following the approach in Section \ref{II} for
computing $S(t)$ by manipulating the rate equation, we provide
a direct method. Let ${\cal E}(t)=S(t)/2$ be the total number of edges/links
in the network at time $t$. Then, a simple rate equation for ${\cal E}(t)$ is:
\begin{equation}\label{12}
    \frac{\partial {\cal E}(t)}{\partial
    t}=m-(c-nc) \left\langle k(t) \right\rangle= m - (c-nc)\frac{S(t)}{N(t)}\ ,
\end{equation}
where the first term is the number of edges brought in by an incoming node, and
the second term is the net number of edges lost due to random deletion of a node.
Substituting $S(t)=2{\cal E}(t)$ and $N(t)=(1-c)t$, we get
\begin{equation}\label{13}
    S(t)=\frac{2m(1-c)}{1+c-2nc}t \;\, \mbox{ and} \;
    \left\langle k(t) \right\rangle= \frac{2m}{1+c-2nc}= \left\langle k \right\rangle_c.
\end{equation}
The validity of Eqn.~(\ref{13}) is checked for different values of
deletion rates, and the results are reported in Fig.~\ref{GROWTH2}.
\\
Inserting $S(t)$ back into Eqn.~(\ref{11}) we get:
\begin{eqnarray}\label{14}
    \frac{\partial k(i,t)}{\partial
    t}&=&\frac{k(i,t)}{2(1-c)t}(1+c-2nc-2c+4nc)\nonumber\\
    &=&\frac{k(i,t)}{2(1-c)t}(1-c+2nc)\ .
\end{eqnarray}
Hence, $\displaystyle k(i,t) = m\left(\frac{t}{i}\right)^\beta$,
where $\displaystyle \beta=\frac{1-c+2nc}{2(1-c)}$. Next, applying Eqn.
(\ref{gam}), we get the power-law exponent to be:
\begin{equation}\label{gam-d}
    \gamma=-1-\frac{2}{1-c+2nc}\ .
\end{equation}
Note that, in this case, there is no singularity when
$c\rightarrow 1$. In fact for $c=1$ and  $0<n\leq n_{\hbox{\small crit}}(1) =1$, we get
\begin{equation}\label{gam-c1}
    |\gamma|=1+\frac{1}{n} \ .
\end{equation}
The magnitudes of the power-law exponents (calculated as the best fit to the cumulative
distribution of the node degrees in simulations) is computed for the
range $c=0\%-90\%$, and the results are checked against
predictions in Fig.(\ref{pl-exp-comp}).


Note that Eqn.~(\ref{13}) is valid only when the denominator is
positive which is equivalent to a finite average degree. So,
$1+c-2nc> 0$, which implies that $-c+2nc< 1$ and
$|\gamma|>1+2/(1+1)=2$. This also implies that for any given $c$,
$\displaystyle 0<n<n_{\hbox{\small crit}}=\frac{1+c}{2c}$. Thus,
{\em for any given deletion rate, $c$, by varying the average
number of compensatory edges for each deleted edge, $n$, one can
program the magnitude of the power-law exponent, $|\gamma|$, to be
anywhere in $(2,\infty)$}. Of course, the price one pays for
getting close to $-2$, is the associated increase in the average
degree, as implied by Eqn.~(\ref{13}). This also, provides a hint
for designing network protocols, that is, too many compensatory
links might make the network unstable.

\subsubsection{The expected degree of a random node at time $t$}
Let's look at the quantity, $E(i,t)$, as defined in
Eqn. (\ref{KD}):
\begin{eqnarray}\label{KD-comp}
    \displaystyle E(i,t)&=&D(i,t)K(i,t)=m\left(\frac{t}{i}\right)^{-\frac{c}{1-c}
    +\beta}\nonumber\\ \displaystyle
    &=&m\left(\frac{t}{i}\right)^{\frac{-2c+(1-c+2nc)}2{1-c}}\nonumber\\ \displaystyle
    &=&m\left(\frac{t}{i}\right)^{\frac{1-c(3-2n)}{2(1-c)}}\ .
\end{eqnarray}
Thus, for $n=1$ the expected degree becomes independent of $c$, and
for and $n>1$ $E(i,t)$ diverges with $t$. Otherwise,
for any $n<1$, if $1\geq c >\frac{1}{3-2n}$, then $E(i,t)\rightarrow 0$
as $\frac{t}{i}\rightarrow \infty$. For example, for $n=3/4$ and $c>2/3$,
the expected degree of a node will decrease with time.
As is well known in the static case, the interesting properties of
scale-free networks are due to the divergence of the second
moment while having a finite mean, which
happens for $|\gamma|<3$. Hence,
an interesting quantity would be:\\
\begin{eqnarray}\label{K2D-comp}
    E_{2}(i,t)&=&D(i,t)K^{2}(i,t)\nonumber\\
    \displaystyle
    &=&m\left(\frac{t}{i}\right)^{\frac{1-2c(1-n)}{(1-c)}}\ .
\end{eqnarray}
So, for any $n>0.5$, and irrespective of the value of $c$,
$E_{2}(i,t)$ diverges, which is consistent with the fact that for
any $n>0.5$, $|\gamma|<3$ and the underlying degree distribution
has unbounded variance. Thus, one might want to work in the
regime, $1>n>0.5$ and $1\geq c >\frac{1}{3-2n}$, where
$E(i,t)\rightarrow 0$ but $E_{2}(i,t) \rightarrow \infty$ as
$\frac{t}{i}\rightarrow \infty$. For example, if $n=0.75$ and
$c\rightarrow 1$, then one can get an exponent of $-2.33$, and yet
have the expected degree of any node to be bounded.

\section{Concluding remarks}\label{V}
We first point out a conceptual link between our
compensatory rewiring scheme discussed in Section \ref{III}, and
the doubly preferential
attachment scheme, as introduced in \cite{Dog3,chung}.
By doubly preferential attachment, we mean that for an edge inserted
in the network, both the initiator and the target nodes are
chosen preferentially based on their degrees. For example, consider
the following random protocol:
At each time step, a new node is inserted that
makes $m$ connections to $m$ preferentially chosen nodes. From the
nodes in the network, $l$ nodes are chosen with probability
proportional to their degrees. Each of these selected nodes
 initiates $m$ new links to $m$ preferentially chosen targets. It can be shown
 \cite{Dog3,chung} that the
power-law exponent $\displaystyle |\gamma|=2+\frac{1}{1+2l}$, and hence, by choosing
$l$
one can make the exponent as close to $-2$ as desired. In this
regard, our compensatory rewiring scheme can be considered as a
natural means for introducing doubly preferential attachments.
By uniformly deleting nodes, a
node looses links with probability proportional to its degree. So
a node initiating a compensatory preferential attachment, intrinsically
introduces doubly preferential attachments. The random deletions of nodes
is thus being used in our stochastic protocol to lead to the emergence of
truly scale-free networks.

One of our main motivations for this work was to design random
protocols that will solve the problem of organizing a
highly-dynamic content sharing network. The first step in this
direction would be to design a local and easily implementable
protocol that would lead to the
emergence of a pre-specified network structure under the usage
constraints imposed by the users. As mentioned in the
introduction, although the network size usually grows for such
networks (more people join such networks), the time scale within
which the size changes is much larger than the time scale within
which the old members of the network log-in and log-off. Hence,
the desired form of the network structure should emerge almost solely due to the
dynamics of the protocol and cannot rely too much
on the growth rate itself. As regarding the desired structure of the network,
motivated by many advantageous aspects of scale free networks, one
might want to come up with protocols that could make the network
self organize into a scale-free structure with a desired power-law
exponent (usually around $-2.5$).

There has been some concern that searches on such power-law
networks might not be scalable; however, our
recent results show that by using bond percolation on the
underlying networks, one can make such networks very efficiently
searchable. In particular, we show that for networks having a power-law degree
distribution with exponent close to $-2$, a traffic efficient
search strategy can be locally implemented. Specifically, we show
that $O(\sqrt{N}log^{2}(N))$ communications on those networks are
sufficient to find each content with probability $1$. This is to
be compared to $\Theta(Nlog(N))$ communications for currently used
broadcast protocols. Also, the search takes only $O(log(N))$ time
steps \cite{scalable}. Thus, scale-free structures with exponents close to $-2$, are not
only observed in current P2P systems, but also are the desirable
structures  for realizing a truly distributed and unstructured P2P
data-base system.

The very high rate of log-offs in real P2P networks prevents the
ordinary preferential attachment scheme from forming a scale free
network, with exponents less than $3$ (as shown in the Sec.
\ref{II}). The local compensation process introduced in Sec.
\ref{III}, however, imposes a scale free structure with an
exponent that can always be kept below $3$. All a node has to do
is to start a new preferential connection, whenever it looses one!
Note that this compensatory procedure is quite natural (and
probably essential) for networks in which the members have to be
part of the giant connected component to be able to have access to
almost all other nodes. In fact, in many clients of the existing
P2P networks, this condition is imposed by always keeping a
constant number of links to active IP addresses. Our numerical
simulations show that graphs resulting from our compensatory
protocol are almost totally connected; that is, a randomly chosen
node with probability one belongs to the giant connected component
of the graph even in the limit of $c=1$. Thus, {\em using our
decentralized compensatory
rewiring protocol one can launch, tune, and maintain a dynamic and
searchable P2P content-sharing system}.

We also believe that our model can, at least intuitively, account
for the degree distributions found in some crawls of P2P networks
like Gnutella. As an example, in \cite{GNUT}, the degree
distribution of the nodes in a crawl of the network was found to
be a power-law with an exponent of $-2.3$. Although Gnutella
protocol \cite{Gnut-prot} does not impose an explicit standard on
how an agent should act when it loses a connection, there are
certain software implementations of Gnutella which try to always
maintain a minimum number of connections by trying to make new
ones when one is lost. Thus, while all clients might not be
compensating for lost edges, it is reasonable to assume that at
least a certain fraction are. As shown in Sec. \ref{III}, if we
pick $n=0.75$ (i.e. 75\% of the lost links are compensated for),
and as $c\rightarrow 1$, the degree distribution is indeed a
power-law with exponent $-2.33$.

To summarize, we have designed truly local and yet universal
protocols which when followed by all nodes result in robust,
totally-connected and scale-free networks with exponents arbitrarily
close to $-2$ even in an \textit{ad hoc}, rapidly changing and
unreliable environment.

\end{document}